\documentclass[%
 aip,
 amsmath,amssymb,
 preprint,%
]{revtex4-1}

\usepackage{graphicx}
\usepackage{dcolumn}
\usepackage{bm}

\usepackage{natbib}
\usepackage[utf8]{inputenc}
\usepackage[T1]{fontenc}
\usepackage{mathptmx}
\usepackage{etoolbox}

\usepackage{color}
\usepackage{ulem}

\makeatletter
\def\@email#1#2{%
 \endgroup
 \patchcmd{\titleblock@produce}
  {\frontmatter@RRAPformat}
  {\frontmatter@RRAPformat{\produce@RRAP{*#1\href{mailto:#2}{#2}}}\frontmatter@RRAPformat}
  {}{}
}%
\makeatother
\begin{document}


\title[Synchronization facilitated by frequency differences]{Synchronization facilitated by frequency differences:\\
Dynamics of coupled-oscillator systems with damaged elements}

\author{Shota Inagawa}
 \affiliation{Department of Science, Kagoshima University.}
 
\author{Hiroki Hata}%
\affiliation{Department of Science, Kagoshima University.
}%

\author{Shigefumi Hata}
 \email{sighata@sci.kagoshima-u.ac.jp}
 \homepage{http://sig-hata.com}
\affiliation{Department of Science, Kagoshima University.
}%

\date{\today}

\begin{abstract}
This study investigates the synchronization dynamics of coupled-oscillator systems in which some of the oscillators are damaged and lose their autonomous oscillations.
The damaged elements are modeled using damped oscillators; thus, the system is composed of both limit-cycle oscillators and damped oscillators.
In this system, as is commonly observed in conventional coupled limit-cycle oscillators, synchronization among oscillators is destroyed when the difference between the natural frequencies of the oscillators increases.
However, in the presence of damped oscillators, synchronization can be facilitated by further increasing the frequency difference from the desynchronization state.
We conduct numerical simulations on coupled Stuart-Landau oscillators and investigate this reentrance of synchronization systematically.
We also propose an approximate theory to predict the stability of the synchronization state based on a linear stability analysis of the fixed point, which reveals the appearance of the Hopf modes.
Using this theory, we argue that the reentrance of synchronization driven by increasing frequency differences can be observed in a wide range of coupled-oscillator systems with damaged elements.

\end{abstract}

\maketitle

\begin{quotation}
Synchronization of rhythmic elements is ubiquitous in nature.
Understanding synchronization mechanisms is important for potential applications in various research fields.
To this end, many theoretical studies have modeled such elements using limit-cycle oscillators and investigated the synchronization dynamics observed in their coupled populations.
It is particularly well known that synchronization is generally inhibited when the difference between the natural frequencies of oscillators is increased.
In this study, we assume a system in which parts of the rhythmic elements can be damaged and lose their autonomous oscillations.
We model such elements using damped oscillators and thus consider a coupled system in which both limit-cycle oscillators and damped oscillators are present.
In this system, as in conventional coupled limit-cycle oscillators, synchronization is destroyed by increasing the frequency difference between the oscillators.
However, in the presence of damped oscillators, synchronization can be facilitated again by further increasing the frequency difference from the desynchronization state.
We conduct numerical simulations on coupled Stuart-Landau oscillators and systematically investigate such a reentrance of synchronization induced by an increase in the frequency difference.
We also propose a theory for predicting the boundary of the synchronization / desynchronization transition in phase diagrams based on a linear stability analysis of the fixed point, and argue that reentrant synchronization can be observed in a wide range of coupled-oscillator systems with damaged elements.
Our results provide a new perspective on the mechanisms of synchronization. 
\end{quotation}

\section{\label{sec:level1}Introduction}
Rhythmic elements are observed at various scales in nature.
Examples include contracting cardiac muscle cells, firing neurons, and turbines in power grids, to name a few.
These microscopic elements interact with each other and form a population to produce macroscopic rhythmic phenomena, such as heartbeats, collective firing of neurons, and synchronized voltage changes between power generators and substations in power grids.
The macroscopic dynamics exhibited by populations of microscopic elements have attracted attention in a wide range of research fields for understanding biological phenomena and engineering applications~\cite{Winfree1980, Glass1988, Pikovsky2001, Manrubia2004, Boccaletti2008, Ermentrout2010, Strogatz1993, Stefanovska1999, Yamaguchi2003, Golombek2010, Holt2016, Asllani2018, Monga2019, Rohden2012, Motter2013, Christensen2009}.
To understand the mechanisms of the macroscopic behavior, theoretical studies have been conducted modeling microscopic elements using limit-cycle oscillators or further simplified phase oscillators, and have investigated the population dynamics observed in their coupled systems, including synchronization~\cite{Winfree1980, Glass1988, Pikovsky2001, Ermentrout2010, Strogatz1993}, clustering~\cite{McGraw2005, Kiss2007, Xie2014, Leon2024}, solitary waves~\cite{Jaros2018, Sathiyadevi2018, Teichmann2019}, chimera~\cite{Kuramoto2002, Abrams2006}, and chaos~\cite{Ku2015, Bick2018}.
Synchronization, where the rhythms of microscopic elements are aligned to produce macroscopic rhythms, has particularly attracted significant attention.

In the real world, an individual rhythmic element may be damaged and lose its autonomous oscillations, such as necrosis in cell populations and mechanical problems in turbines in the power grid.
Previous studies have modeled such inactive elements using damped oscillators and excitable media, and investigated the effects of such microscopic defects on the entire system.
They revealed that, in a coupled system of active limit-cycle oscillators and inactive damped oscillators, all elements stop oscillating and macroscopic oscillations disappear when the number of inactive oscillators exceed a certain threshold, even if active oscillators remain in the system (aging transition)~\cite{Daido2004, Daido2016}.
The aging transition was first observed in globally coupled Stuart-Landau oscillators~\cite{Daido2004} and was also found in random networks~\cite{Morino2011, Tanaka2012}.
This transition has also been observed when inactive elements are modeled as excitable media~\cite{Daido2016} and has
subsequently been studied in coupled neuron models~\cite{Biswas2022, Sharma2025}.
These results indicate that microscopic damage can have a significant impact on macroscopic population dynamics in coupled-oscillator systems.
However, the effects of such microscopic defects on the synchronization dynamics are not yet fully understood.

In this study, we investigate the synchronization dynamics observed in coupled-oscillator systems in which both limit-cycle oscillators and damped oscillators are present.
We employ Stuart-Landau oscillators to describe both types of oscillators.
We consider a globally-coupled oscillator population composed of three types of oscillators: two types of limit-cycle oscillators with different natural frequencies and one type of damped oscillator.
As in previous studies, we assume that the oscillators synchronize completely in each subpopulation and take a continuous limit to describe the system dynamics using three coupled differential equations.
In this system, synchronization among oscillators is destroyed when the difference between the natural frequencies of the oscillators increases, as in conventional coupled limit-cycle oscillators.
However, in the presence of damped oscillators, synchronization can be facilitated by further increasing the frequency difference from the desynchronization state (reentrant synchronization).
We conduct numerical simulations to systematically investigate the conditions under which the reentrance of synchronization is observed in the system.
We also propose an approximate theory to predict the stability of the synchronization state based on a linear stability analysis of the fixed point, which reveals the appearance of the Hopf modes.
Using the proposed theory, we discuss how reentrant synchronization can be observed in a wide range of coupled-oscillator systems with damped oscillators.

This paper is organized as follows.
In Sec. II, we formulate a mathematical model using Stuart-Landau oscillators, where both limit-cycle and damped oscillators are present.
In Sec. III, we present the typical dynamics observed in the system and the phase diagrams for three different states: synchronization, desynchronization, and amplitude death.
In Sec. IV, we propose a theory that predicts the boundaries of the three states in the phase diagrams based on a linear stability analysis of the fixed point.
In Sec. V, we investigate the shapes of the phase boundaries based on the preceding theory.
Finally, we summarize the results and discuss future studies in Sec. VI.

\section{\label{sec:level2}Model description}
We consider globally coupled Stuart-Landau oscillators of size $N$:
\begin{align}
\frac{\textrm d z_j}{\textrm d t}=
\left ( \alpha_j + i \omega_j \right ) z_j - \left ( \beta_j+i \gamma_j \right ) \left | z_j \right |^2 z_j + \frac{K}{N}\sum_{k=1}^N \left (z_k - z_j \right ),
\label{eq01}
\end{align}
where $z_j \in \mathbb C$ denotes the complex amplitude of the $j$-th oscillator. 
The coupling strength is specified by $K$.
Parameter $\alpha_j$ encodes the strength of the active oscillations.
When isolated, that is $K = 0$, each oscillator exhibits a limit-cycle oscillation ($\alpha_j > 0$) or damped oscillation ($\alpha_j < 0$) depending on the sign of $\alpha_j$.
The parameters $\beta_j>0$ and $\gamma_j$ measure the nonlinearity of amplitude and phase, respectively.
The natural frequency of the oscillator is $\omega_j - \alpha_j \gamma_j / \beta_j$.

We consider that the oscillator population is composed of three types of oscillators: two types of active oscillators that exhibit limit-cycle oscillation when isolated and inactive oscillators that exhibit damped oscillation when isolated.
We assume that the two types of active oscillators differ only in the frequency parameter $\omega_j$ and share the other parameters $\alpha_j = \alpha_A>0, \beta_j = \beta_A$ and $\gamma_j = \gamma_A$.
We denote the population ratio of the inactive oscillator by $p \in [0, 1]$, that is, the number is $Np$, and we assume that the ratios of the two active oscillators are equivalent,
such that the population ratio of the three oscillators is $(1-p)/2:(1-p)/2:p$.
We also assume that the oscillators synchronize completely in each subpopulation, that is, the three types of oscillators share their own value of $z_j$ in the subpopulation, which allows us to denote the state as $A_1, A_2$ for the active oscillators and $I$ for the inactive oscillators.
We choose the appropriate scale and measuring frame to rescale the variables and parameters as $z_j \rightarrow \sqrt{\alpha_A/\beta_A}z_j, t\rightarrow t / \alpha_A, \omega_j \rightarrow \alpha_A \omega_j, \gamma_j \rightarrow \beta_A \gamma_j$ and $K \rightarrow \alpha_A K$.
Taking the limit $N \rightarrow \infty$, the model can be reduced to the following coupled differential equations:
\begin{align}
\frac{\textrm d A_1}{\textrm d t}
= & \left ( 1 + i \omega_{A1} \right ) A_1 - \left ( 1+i \gamma_{A} \right ) \left | A_1 \right |^2 A_1\cr
& + K\left \{ p\left (I - A_1 \right ) + \frac{1-p}{2}\left (A_2 - A_1 \right ) \right \},
\label{eq02}\\
\frac{\textrm d A_2}{\textrm d t}
= & \left ( 1 + i \omega_{A2} \right ) A_2 - \left ( 1+i \gamma_{A} \right ) \left | A_2 \right |^2 A_2\cr
& + K\left \{ p\left (I - A_2 \right ) + \frac{1-p}{2}\left (A_1 - A_2 \right ) \right \},
\label{eq03}\\
\frac{\textrm d I}{\textrm d t}
= & \left ( -\alpha + i \omega_I \right ) I - \left ( \beta +i \gamma_I \right ) \left | I \right |^2 I\cr
& + K\left \{ \frac{1-p}{2}\left (A_1 - I \right ) + \frac{1-p}{2}\left (A_2 - I \right ) \right \},
\label{eq04}
\end{align}
where the parameters indexed by $A1, A2$ and $I$ are shared by each subpopulation.
The parameters $\alpha = \left | \alpha_I / \alpha_A \right |>0$ and $\beta = \beta_I / \beta_A$ in Eq.~(\ref{eq04}) quantify the ratio of the parameters in the active and inactive oscillators. 
In the main text, we set $\beta = 1$ and $\gamma_A = \gamma_I = 0$ for simplicity; thus, the natural frequency of each subpopulation is given by $\omega_{A1}, \omega_{A2}$ and $\omega_{I}$.
The results for $\beta \neq 1, \gamma_A = \gamma_I \neq 0$ are presented in the Appendix.

\section{\label{sec:level3}Synchronization dynamics}
First, we numerically investigate the typical dynamics observed in Eqs.~(\ref{eq02})-(\ref{eq04}).
The model exhibits three different states:
synchronization, desynchronization, and amplitude death.
Figure~\ref{fig01}(a) shows the phase diagram of these three states depending on the frequency difference $\Delta \omega = \omega_{A2} - \omega_{A1}$ of the active oscillators and the population ratio $p$ of the inactive oscillator.
The natural frequencies are set to $\omega_{A1} = 5.0, \omega_{A2} = 5.0+\Delta \omega$ and $\omega_{I}= 5.0$.
The damping parameter is set to $\alpha = 1.0$.
The coupling constant is set to $K=1.2$.
Now, we examine the transition in the population dynamics of the oscillators by gradually increasing the frequency difference from $\Delta \omega = 0$.
The diagram shows that an increase in the frequency difference $\left |\Delta \omega \right |$ destroys synchrony when the population ratio $p$ of the inactive oscillators is relatively small, for example $p \simeq 0.65$, which is typically observed in various coupled-oscillator systems.
When the ratio $p$ is large, for example $p \simeq 0.85$, an increase in $\Delta \omega$ does not lead to desynchronization, and instead causes a decrease in the oscillation amplitude, leading to amplitude death, known as the aging transition~\cite{Daido2004}.

When the ratio is between the above two situations ($p \simeq 0.7$), the system exhibits intriguing behavior.
If the frequency difference increases from $\Delta \omega = 0$, synchronization is destroyed (see time series of $\Delta \omega = 0.4 \rightarrow 2.0$).
However, if $\Delta \omega$ is increased further, the synchronized state recovers ($\Delta \omega = 2.0 \rightarrow 3.2$).
Thus, an increase in the frequency difference $\Delta \omega$ can facilitate the synchronization of the oscillator population and induces reentrance of the system into the synchronization region.
Figure~\ref{fig01}(b) and (c) show the phase diagrams for $\omega_{A2} = 4.0$ and $6.0$, respectively.
The other parameters are set to be the same as those used in panel (a).
In both panels, we can see the reentrant synchronization regions ($p\simeq 0.72$, $\Delta \omega \simeq 3$ for (b), and $\Delta \omega \simeq -3$ for (c)).
Thus, the reentrance of synchronization is observed with different parameter values.
This peculiar transition is not observed in the population of general limit-cycle oscillators, and therefore, it would be typical behavior in the presence of damped oscillators.

\section{\label{sec:level4}Theoretical prediction of transition boundaries}
In this section, we propose a theory for predicting the boundaries of the synchronization and desynchronization regions in the phase diagrams.
The transition of synchronization / desynchronization can generally be predicted by conducting a linear stability analysis of phase differences.
The frequency synchronization of the two oscillators is characterized by the convergence of the phase difference to a fixed value.
If the phase difference does not converge and continues to increase (or decrease) over time, the phases of the two oscillators are {\it slipping}, indicating that the system is in a desynchronization state.
In particular, for diffusively coupled SL oscillators, stability analysis can be performed analytically~\cite{Aronson1990, Matthews1990}.

In this study, we instead employ a linear stability analysis for the fixed point $(A_1, A_2, I) = (0, 0, 0)$ of the system~(\ref{eq02})–(\ref{eq04}) in $\mathbb C^3$ space to evaluate the stability of the synchronization state.
This analysis is typically used to predict the amplitude death in coupled SL systems.
Amplitude death is characterized by a stable fixed point because the system converges to a fixed point and does not exhibit oscillation.
If the fixed point becomes unstable, the system deviates from the fixed point via Hopf bifurcation and converges to a stable limit cycle, which corresponds to a stable synchronous oscillation of the entire system.
We then focus on the excitation of the second Hopf mode to predict the synchronization-desynchronization boundary.
As previously noted, the appearance of the first Hopf mode induces a stable synchronous oscillation in the coupled SL system.
The second mode is then expected to disturb the synchronized oscillation by injecting another Hopf frequency, leading the system to a desynchronization state.
In diffusively coupled SL systems, a phase transition from amplitude death to desynchronization can be observed when two Hopf modes are simultaneously excited ~\cite{Aronson1990}.
This suggests that the second Hopf mode determines the stability of the synchronous oscillations.
Note that the system deviates from the fixed point after the excitation of the first Hopf mode, and the stability of the second Hopf mode cannot generally be evaluated by the analysis of the fixed point, but rather requires analysis of the limit cycle.
However, in the current study, we focus on the dynamics when the ratio $p$ of inactive oscillators is relatively large.
In such a situation, the oscillatory amplitude of the entire system is small owing to the presence of inactive units, and the growth rate of the second Hopf mode can be approximately evaluated by that of the fixed point.
In Fig.~\ref{fig01} we plotted the border curves where Hopf modes appeared at a fixed point.
These are in good agreement with the boundaries of the three states.
Thus, we employed linear stability analysis on the fixed point to predict the appearance of the second Hopf mode, which governs the synchronization / desynchronization transition.

\section{\label{sec:level5}Shape of boundaries: Generality of reentrant synchrony}
In this section, the shapes of the border curves for the three states are investigated to confirm the generality of reentrant synchronization.
For mathematical tractability, we denote the natural frequency of each subpopulation as $\omega_{A1} = \omega_1 + \Delta \Omega, \omega_{A2} = \omega_1 - \Delta \Omega$ and $\omega_I = \omega_2$.
Note that we can represent any set of $\omega_{A1}, \omega_{A2}$ and $\omega_I$ by changing the parameters $\omega_1, \omega_2$ and $\Delta \Omega$.
First, we set $\omega_1 = \omega_2$, that is, the frequencies of the active oscillators $\omega_{A1}$ and $\omega_{A2}$ are symmetrical with respect to $\omega_I$.
This assumption allows us to calculate the border curves analytically.
We then relax the above assumptions and investigate how reentrant synchrony regions are formed in the phase diagrams.

A linear stability analysis is performed by linearizing Eqs.(\ref{eq02})-(\ref{eq04}) around the fixed point $(A_1, A_2, I)=(0, 0, 0)$.
The system has three Hopf modes, and the growth rate $\lambda$ of each mode is obtained by this analysis, which is determined by the cubic equation:
\begin{align}
\lambda^3 + a \lambda^2 + b \lambda + c =0,
\label{eq00}
\end{align}
where the coefficients $a,b$ and $c$ are determined by the parameters $\alpha, \omega_1, \omega_2, K, p$ and $\Delta \Omega$.
The coefficients do not depend on the nonlinearity parameters $\beta, \gamma_A$ and $\gamma_I$ which disappear in the linearized equation.
The stability of each Hopf mode is determined by the sign of $\textrm{Re} \lambda$. Thus, the border curves are characterized by $\textrm{Re} \lambda = 0$.
When $\omega_1 = \omega_2$, the coefficients $a,b$ and $c$ in Eq.~(\ref{eq00}) are real numbers.
The complex conjugate theorem states that a cubic equation with real coefficients has either three real roots or one real root and a pair of complex conjugate roots.
Therefore, the cases in which Hopf modes become excited are limited to the following two cases:
(i) A single real root $\lambda$ becomes zero, or (ii) the real parts of complex conjugate roots $\lambda, \lambda^\dagger$ become zero simultaneously.
This limitation allows us to obtain the border curves analytically as functions of $\Delta \Omega$ (see Appendix for the detailed derivation):
\begin{align}
p_{\textrm i}(\Delta \Omega) = &
\frac{\left ({\Delta \Omega}^2 + 1 -K   \right )(K+\alpha)}{K \left \{ {\Delta \Omega}^2 + (1-K)(1 + \alpha ) \right \}},
\label{eq05}\\
p_{\textrm{ii}}(\Delta \Omega) = &
 \frac{\left \{{\Delta \Omega}^2 + 2K^2+ (1-\alpha)(1- 3K + \alpha) \right \}(2-K)}{K \left \{{\Delta \Omega}^2 - (1 + \alpha)(1+K-\alpha) \right \}}.
\label{eq06}
\end{align}
Note that an additional condition is required for Hopf modes to be excited on the curve $p_{\textrm{ii}}$ that the cubic equation~(\ref{eq00}) has complex conjugate roots, which will be discussed later.

Hereafter, we restrict the coupling constant to $K\in(1,2)$ such that the secondary curve has the ratio $p_{\textrm{ii}} \in (0, 1)$.
Figure~\ref{fig02}(a) shows an example of curves $p_{\textrm{i}}$ and $p_{\textrm{ii}}$.
It is obvious from Eqs.(\ref{eq05}) and (\ref{eq06}) that both $p_{\textrm{i}}$ and $p_{\textrm{ii}}$ are even functions of $\Delta \Omega$, and we present them only for $\Delta \Omega \geq 0$.
We can check Eqs.(\ref{eq05}) and (\ref{eq06}) to confirm that the two curves $p_{\textrm{i}}$ and $p_{\textrm{ii}}$ satisfy 
\begin{align}
& p_{\textrm{i}}(0)>p_{\textrm{ii}}(0),
\label{eq07}\\
& \lim_{{\Delta \Omega}^2\rightarrow (K-1)(\alpha + 1 )\pm 0} p_{\textrm{i}}(\Delta \Omega)=\pm\infty,
\label{eq08}\\
&
\lim_{|\Delta \Omega| \rightarrow \infty} p_{\textrm{i}}(\Delta \Omega) = \frac{K + \alpha}{K}>1,
\label{eq085}\\
& \lim_{|\Delta \Omega| \rightarrow \infty} p_{\textrm{ii}}(\Delta \Omega) = \frac{2-K}{K}\in(0,1).
\label{eq09}
\end{align}
The evaluation of the value range in Eq.~(\ref{eq09}) comes from $K\in(1,2)$.
The equations also show that both $p_{\textrm{i}}$ and $p_{\textrm{ii}}$ are decreasing functions of $|\Delta \Omega|$.
Based on these facts, we can confirm that the shapes of $p_{\textrm{i}}$ and $p_{\textrm{ii}}$ are qualitatively identical to those in Fig.~\ref{fig02}(a) with the intersection $(p^*, \Delta \Omega^*)$ in $p\in (0,1)$.

Let us now check whether the cubic equation~(\ref{eq00}) has complex conjugate roots on the curve $p_{\textrm{ii}}$.
This is verified by calculating the discriminant $D = a^2 b^2 - 4b^3 - 4 a^3 c +18 abc - 27 c^2$.
We can analytically obtain that $D<0$ holds for $p_{\textrm{ii}}$ if $|\Delta \Omega|<\Delta \Omega^*$, that is, $\Delta \Omega \in (-\Delta \Omega^*, \Delta \Omega^*)$ (see the Appendix for details).
This indicates that the growth rates $\lambda$ are not complex conjugates on $p_{\textrm{ii}}$ within this interval.
Consequently, the Hopf modes are not excited.
We compare the analytical investigations and the numerical results in Fig.~\ref{fig02}(b).
As can be seen, the analytical results agree well with the numerical results.

Finally, we relaxed the symmetric constraint $\omega_1 = \omega_2$ in frequencies to observe the formation of reentrant synchronization region in the phase diagram (Figure~\ref{fig03}).
The three panels show the phase diagrams for $\omega_2 = 5.0, 5.25$, and $5.5$.
The boundary curves in the latter two panels are obtained by solving Eq.~(\ref{eq00}) numerically.
The removal of the symmetric constraint causes the cubic equation to have coefficients of complex numbers and no longer have complex conjugate roots.
The two Hopf modes that become excited at $p_{\textrm{ii}}$ in the symmetric case are excited independently.
Thus, we see that $p_{\textrm{ii}}$ observed in the symmetric case splits into two boundaries that are smoothly connected to $p_{\textrm i}$, allowing the synchronization region to appear between them.
Note that because $p_{\textrm{ii}}$ is a decreasing function of $|\Delta \Omega|$,
In the vicinity of the breaking of symmetry in the frequencies, the slopes of the boundaries that were originally $p_{\textrm{ii}}$ are negative.
This guarantees that a reentrant synchronization region is always formed in the system.

\section{\label{sec:level6}Conclusions and Discussion}
In this study, we investigated the synchronization dynamics of coupled-oscillator systems with damaged elements.
The damaged elements were modeled using damped oscillators.
As we have shown, the presence of damaged elements may play a decisive role in the dynamics of the system:
when the frequency difference between the active elements increases from the desynchronization state,
the stability of the synchronization state can be facilitated, leading the system to exhibit a reentrant transition to synchronization.
We propose a theory to account for the above observations based on a linear stability analysis of a fixed point.
This theory accurately predicts the boundaries of the synchronization and desynchronization states in phase diagrams.

We emphasize that the presence of damaged elements can facilitate the synchronization of all oscillators in a wide range of coupled-oscillator systems.
We conducted theoretical investigations by employing the SL oscillator, which is the normal mode of the supercritical Hopf bifurcation.
This indicates that our theory may hold for any oscillator population with damaged elements where each model oscillator is close to the supercritical Hopf bifurcation.
Our analytical investigations suggest that when such a system has a certain symmetry in the frequency distribution, there is always a boundary curve in the phase diagram where two Hopf modes are excited simultaneously.
This boundary splits into two when symmetry is broken, and a synchronous region appears between them.
Importantly, the shape of this boundary is qualitatively the same irrespective of the parameter values, which guarantees the formation of reentrant synchronization regions in the phase diagrams.

The results of this study suggest that the presence of damaged elements may help to synchronize the entire system.
Specifically, the population ratio of such elements can be a control parameter for synchronization, as shown in the phase diagrams.
This study may pave the way for a new scenario for synchronization and has the potential to provide new insights into applications.

\section*{Acknowledgment}
The authors acknowledge H. Nakao, R. Muolo and I. Le\'{o}n for the useful discussions.
S.H. acknowledges JSPS KAKENHI 25K15264 for financial support.

\section*{Data availability}
The data that support the findings of this study are available within the article.

\appendix
\section{Linear stability analysis of the system}
The linearized equation of Eqs~(\ref{eq02})-(\ref{eq04}) is given by
\begin{align}
\frac{\textrm d}{\textrm d t}\vec X = \hat J \vec X,
\label{eqa01}
\end{align}
where $\vec X = {}^t(\delta A_1, \delta A_2, \delta I)$ is the state vector representing the perturbative deviation from the fixed point.
The Jacobian matrix $\hat J$ for a fixed point $(A_1, A_2, I) = (0,0,0)$ is given by
\begin{align}
\hat J =
\begin{pmatrix}
1+i \omega_{A1}-K\frac{p+1}{2} & Kp & K\frac{1-p}{2}\\
K\frac{1-p}{2} & 1+i \omega_{A2}-K\frac{p+1}{2} & Kp\\
K\frac{1-p}{2} & K\frac{1-p}{2} & \alpha+i \omega_{I}-K(1-p)
\end{pmatrix}
.
\label{eqa02}
\end{align}
The linear growth rates $\lambda$ of each eigenmode (Hopf mode) are then determined using the characteristic polynomial
\begin{align}
\textrm{det} \left ( \lambda \hat I - \hat J \right ) =0,
\label{eqa03}
\end{align}
where $\hat I$ denotes an identity matrix.

\section{Derivation of boundary curves $p_{\textrm i}$ and $p_{\textrm{ii}}$}
We assume that $\omega_{A1,2} = \omega_1 \pm \Delta \Omega$, $\omega_I = \omega_2 = \omega_1$.
Because of the rotational symmetry of Stuart-Landau oscillators, we can set $\omega_1 = 0$ without loss of generality.
The characteristic polynomial~(\ref{eqa02}) is expressed as Eq.~(\ref{eq00}) with coefficients
\begin{align}
& a = 2K -2 + \alpha,
\label{eqa04}\\
& b = K^2 + \left \{ ( 1 + \alpha) p - 3 + \alpha \right \} K + {\Delta \Omega}^2 + 1 -2 \alpha,
\label{eqa05}\\
& c = \left \{ (1 + \alpha) p -\alpha \right \} K^2 + \left \{ ({\Delta \Omega}^2 + 1) (1-p) - \alpha (1+p) \right \} K + \alpha {\Delta \Omega}^2 +  \alpha.
\label{eqa06}
\end{align}
Note that the coefficients $a,b$ and $c$ are real numbers.

The complex conjugate theorem states that a cubic equation with real coefficients has either three real roots or one real root and a pair of complex conjugate roots.
In the former case, the border curves are characterized by $\lambda = 0$, which yields $c=0$.
We would have $c=0$ when the real root takes $\lambda = 0$ also in the latter case.
Then, the border curve $p_{\textrm i}$ is given by $c=0$.
When the real part of the complex conjugate roots is $\textrm{Re} \lambda = \textrm{Re} \lambda^\dagger = 0$, we obtain the condition $c = ab$, which yields the border curve $p_{\textrm{ii}}$.

\section{Existence of complex conjugate roots on $p_{\textrm{ii}}$}
A cubic equation with real coefficients has complex conjugate roots when the discriminant $D = a^2 b^2 - 4b^3 - 4 a^3 c +18 abc - 27 c^2$ is positive.
In the present case, we substitute the coefficients~(\ref{eqa04})-(\ref{eqa06}) and $p=p_{\textrm{ii}}$ to obtain:
\begin{align}
& D = - \frac{4f^2 g}{h^3},
\label{eqa07}\\
& f = {\Delta \Omega}^4 + \left \{ K(5 K -11 + 5\alpha) + 6-4 \alpha +2 {\alpha}^2 \right \}{\Delta \Omega}^2 +(K-1+\alpha)^2(3K-3+\alpha)(1+\alpha),
\label{eqa08}\\
& g = {\Delta \Omega}^4 + \left \{K^2 -(3 - \alpha)K+2 +{\alpha}^2\right \} {\Delta \Omega}^2 -(K-1)^2(K-1 + \alpha)(1 + \alpha),
\label{eqa09}\\
& h = -(K-1+\alpha)(1 + \alpha) - {\Delta \Omega}^2.
\label{eqa10}
\end{align}
We have $h<0$ since $K \in (1 ,2)$.
$g=g({\Delta \Omega}^2)$ is a convex function of ${\Delta \Omega}^2$, and it has been verified that it satisfies
$g(0)= -(K-1)^2(K-1 + \alpha)(1 + \alpha_I) <0$ and $g({\Delta \Omega^*}^2) = 0$,
where ${\Delta \Omega^*}^2$ is obtained by $p_{\textrm i}({\Delta \Omega^*}^2)=p_{\textrm{ii}}({\Delta \Omega^*}^2)$ as follows:
\begin{align}
& {\Delta \Omega^*}^2 = \frac{-K^2 + (3 - \alpha) K - 2 - {\alpha}^2+\sqrt{K^4 -2(1-3\alpha)K^3 +(1-14\alpha +7{\alpha}^2)K^2}}{2}.
\label{eqa11}
\end{align}
This implies that $g<0$ holds for $|\Delta \Omega| < \Delta \Omega^*$.
Thus, we have $D < 0$ in $|\Delta \Omega| < \Delta \Omega^*$, indicating that the cubic equation~(\ref{eq00}) does not have complex conjugate roots on $p_{\textrm{ii}}$ within this range.

\section{Nonlinearity parameters $\beta_j$ and $\gamma_j$}
\begin{figure*}
\centering
\includegraphics[width= 0.97 \hsize]{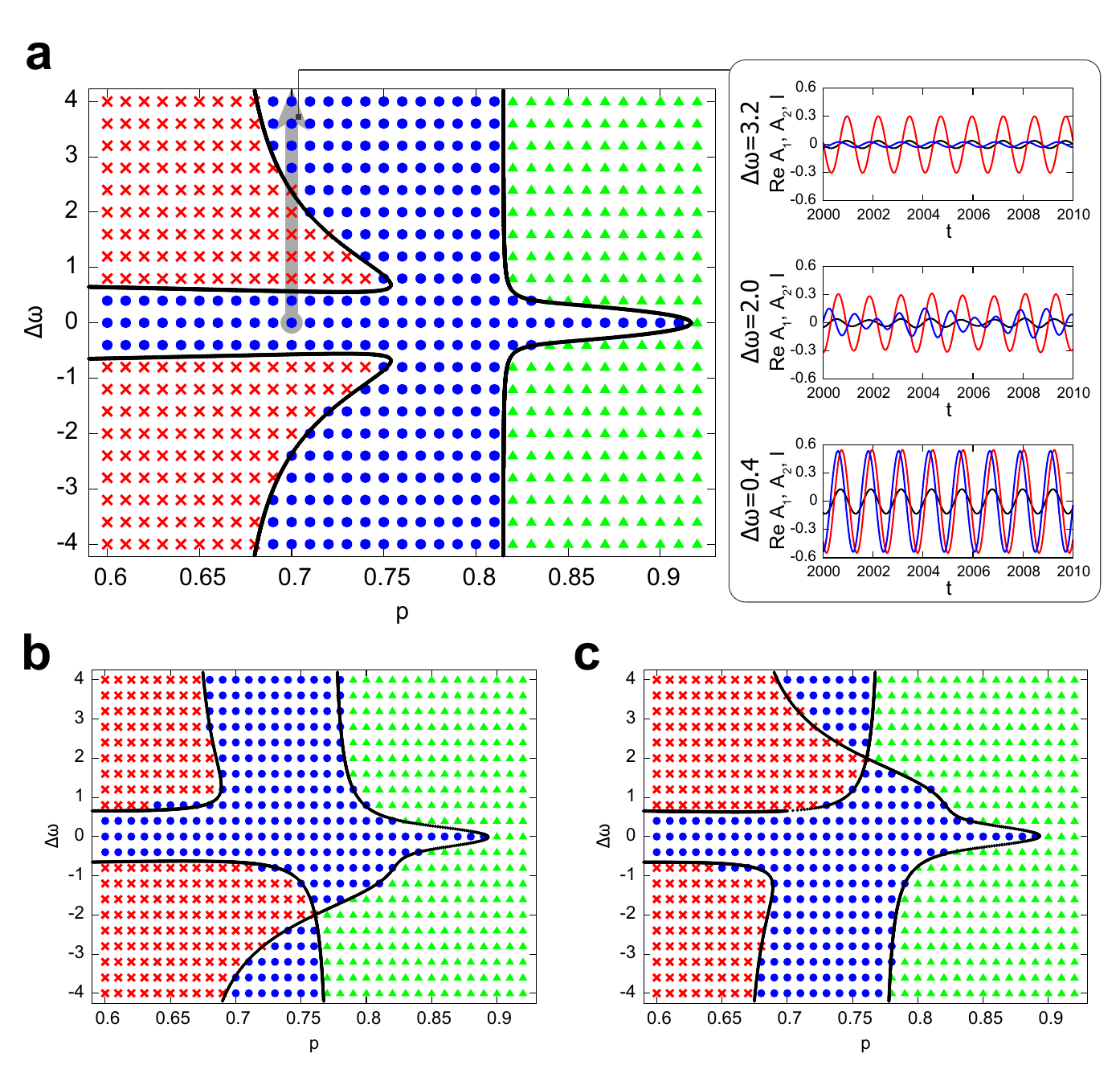}
\caption{\label{fig01}
Phase diagram for synchronization, desynchronization, and amplitude death depending on frequency difference $\Delta \omega = \omega_{A2} - \omega_{A1}$ and the population ratio $p$ of inactive oscillator.
Blue circles, red crosses, and green triangles plot respectively the parameter values of synchronization, desynchronization, and amplitude death identified from numerical simulations.
Black curves indicate theoretical predictions of transition boundaries (see Section IV).
(a) $\omega_{A1} = 5.0, \omega_{A2} = 5.0 + \Delta \omega$ and $\omega_{I} = 5.0$.
The time series of each oscillator for $p=0.7$ are shown in the insets to illustrate the typical dynamics.
Real part of $A_1, A_2$ and $I$ are plotted respectively by blue, red, and black curves.
Frequency differences are respectively set as $\Delta \omega = 0.4, 2.0$ and $3.2$.
Parameter changes between them are highlighted in the phase diagram by a gray arrow.
(b) $\omega_{A1} = 5.0, \omega_{A2} = 5.0 + \Delta \omega$ and $\omega_{I} = 4.0$.
(c) $\omega_{A1} = 5.0, \omega_{A2} = 5.0 + \Delta \omega$ and $\omega_{I} = 6.0$.
Other parameters are set as $\alpha = 1.0, \beta = 1.0$ and $\gamma_A = \gamma_I = 0$ for all panels.
}
\end{figure*}

\begin{figure*}
\centering
\includegraphics[width= 0.9 \hsize]{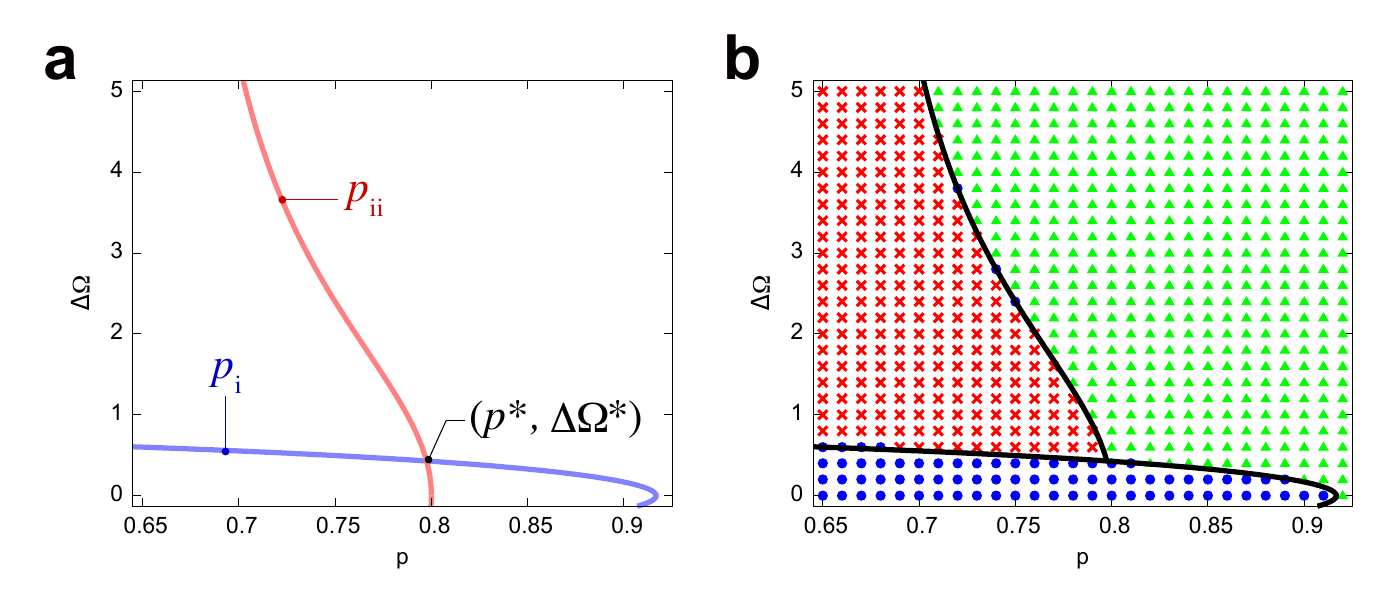}
\caption{\label{fig02}
(a) Border curves $p_{\textrm{i}}(\Delta \Omega)$ and $p_{\textrm{ii}}(\Delta \Omega)$.
(b) Comparison of analytical and numerical results.
Blue circles, red crosses, and green triangles plot respectively the parameter values of synchronization, desynchronization, and amplitude death identified from numerical simulations.
Black curves indicate the theoretical prediction of transition boundaries obtained by solving a cubic equation~(\ref{eq00}).
Parameters are set as $\omega_1 = \omega_2 = 5.0, K = 1.2, \alpha = 1.0, \beta = 1$ and $\gamma_A = \gamma_I = 0$.
}
\end{figure*}

\begin{figure*}
\centering
\includegraphics[width= 1.0 \hsize]{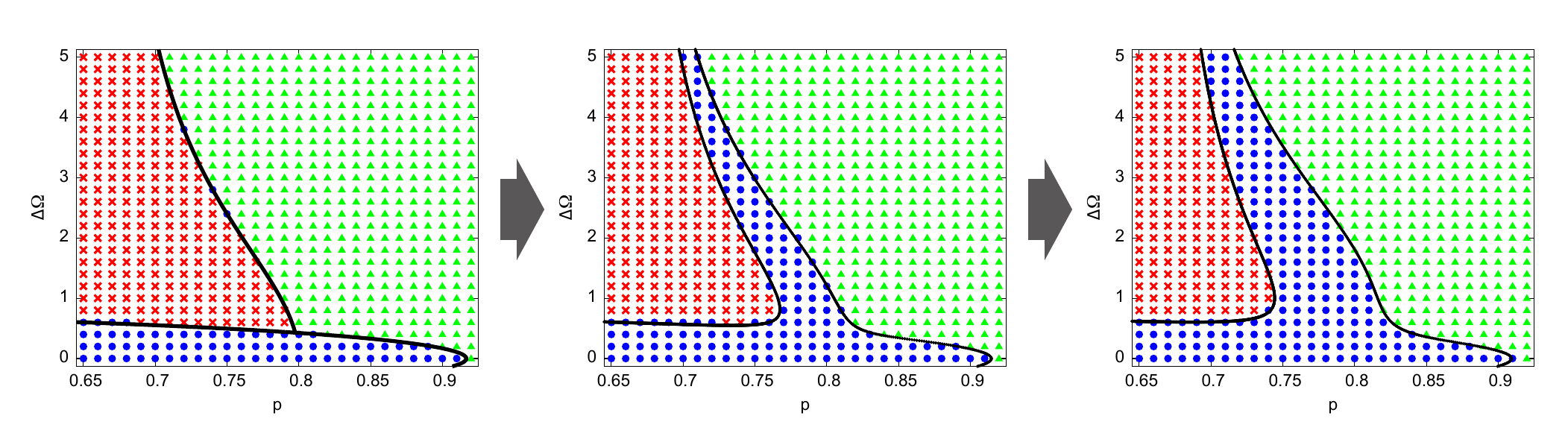}
\caption{\label{fig03}
Formation of the reentrant synchronization region in the phase diagram by breaking the symmetry in the frequency distribution.
Blue circles, red crosses, and green triangles plot respectively the parameter values of synchronization, desynchronization, and amplitude death identified from numerical simulations.
Black curves indicate the theoretical prediction of transition boundaries obtained by solving a cubic equation~(\ref{eq00}).
Border curves are obtained by solving the cubic equation numerically.
Parameters are set as $\omega_1 = 5.0, K = 1.2, \alpha = 1, \beta=1$ and $\gamma_A = \gamma_I = 0$ for all panels.
Results for (left) $\omega_2 = 5.0$, (center) $\omega_2 = 5.25$ and (right) $\omega_2 = 5.5$.
}
\end{figure*}

\begin{figure}
\centering
\includegraphics[width= 0.45 \hsize]{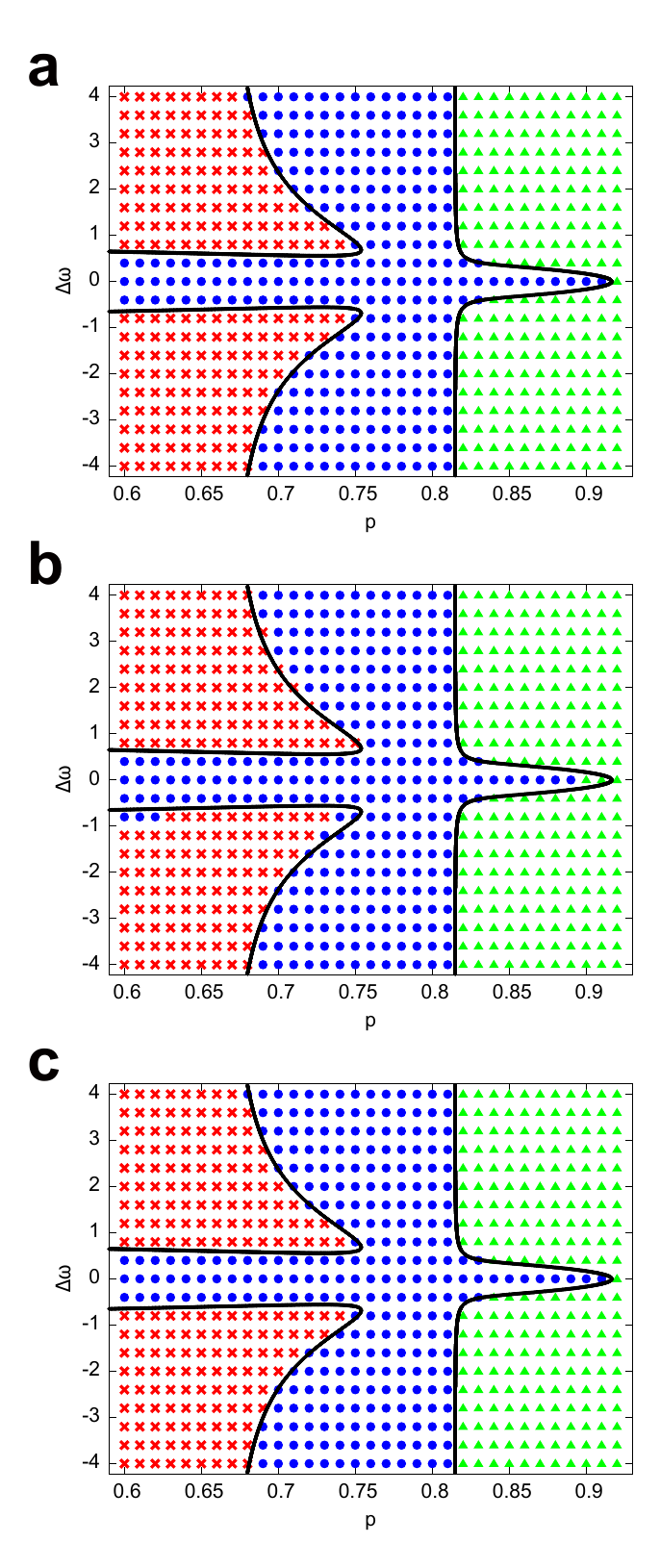}
\caption{\label{fig04}
Phase diagrams for (a)$(\beta, \gamma_A, \gamma_I) = (2.0, 0, 0)$, (b)$(\beta, \gamma_A, \gamma_I) = (1.0, 1.0, 1.0)$ and (c)$(\beta, \gamma_A, \gamma_I) = (2.0, 1.0, 1.0)$.
Blue circles, red crosses, and green triangles plot respectively the parameter values of synchronization, desynchronization, and amplitude death identified from numerical simulations.
Black curves indicate the theoretical prediction of transition boundaries obtained by solving a cubic equation~(\ref{eq00}).
Other parameters are set as $\omega_{A1} = 5.0, \omega_{A2} = 5.0 + \Delta \omega$, $\omega_{I} = 5.0$, $\alpha = 1.0$ and $K=1.2$.
}
\end{figure}

For simplicity, we set the parameters in nonlinear terms as $\beta=1$ and $\gamma_A = \gamma_I = 0$ in the main text.
This section discusses the effects of these parameters.
In the present case, because the parameters $\beta_j$ and  $\gamma_j$ are in the cubic term of $z_j$, they do not appear in the linear stability analysis for the fixed point $(A_1, A_2, I)=(0,0,0)$ (see Eq.~(\ref{eqa02})).
Thus, our theory indicates that these parameters do not affect the phase diagrams at the first order of $\vec X$ in Eq.~(\ref{eqa01}).
This is confirmed by comparing numerical results and theoretical predictions.
Figure~\ref{fig04} shows the results for several values of $(\beta, \gamma_A, \gamma_I)$.
The other parameters are set as shown in Fig.~\ref{fig01}(a).
As can be seen in all panels, the phase diagrams obtained by the numerical simulations are almost the same. This indicates that the parameters $(\beta, \gamma_A, \gamma_I)$ do not significantly affect the results presented in the main text, which are also predicted by the proposed theory.

\bibliography{DamagedSync}

\newpage

\end{document}